\documentclass[10pt,twocolumn,letterpaper]{article}

\usepackage[pagenumbers]{iccv} 

\definecolor{iccvblue}{rgb}{0.21,0.49,0.74}

\usepackage[pagebackref,breaklinks,colorlinks,allcolors=iccvblue]{hyperref}
\usepackage{tikz}
\usepackage[most]{tcolorbox}
\usepackage{svg}
\usepackage{soul}
\usepackage{wrapfig}
\usepackage{bm}
\usepackage{adjustbox}
\usepackage{pifont}
\usepackage{graphicx}
\usepackage{amssymb}
\usepackage{booktabs}
\usepackage{makecell} 
\usepackage{multirow}
\usepackage{multicol}
\definecolor{Gray}{gray}{0.9}
\definecolor{myblue}{RGB}{230, 230, 248}
\definecolor{featherbrown}{HTML}{8C564B}
\definecolor{featherbackgroundbrown}{HTML}{F1E1DB}
\definecolor{fastvgreen}{HTML}{3B7D23}
\definecolor{fastvbackgroundgreen}{HTML}{EDF8E7}
\definecolor{localizationpurple}{HTML}{78206E}
\definecolor{localizationbackgroundpurple}{HTML}{F8E7F6}
\usepackage{amsmath}
\usepackage{enumitem}
\usepackage{algorithm}
\usepackage{algpseudocode}

\usepackage{multirow} 
\usepackage{multicol}
\usepackage[most]{tcolorbox}
\usepackage{listings}
\usepackage{balance}

\definecolor{colorcommentbg}{HTML}{CCCCFF}
\definecolor{colorcommentframe}{HTML}{76608A}
\newcounter{reviewcomment@counter}[section]
\newenvironment{contentagnosticlayout}[1][]{\refstepcounter{reviewcomment@counter}
	\begin{tcolorbox}[adjusted title={Q-Insight Prompt}, fonttitle={\bfseries\footnotesize}, fontupper=\footnotesize, colback={colorcommentbg!40}, colframe={colorcommentframe!90},coltitle={white},#1]
}{\end{tcolorbox}}

\usepackage{colortbl}
\hypersetup{
	colorlinks=true,
	citecolor=teal,
}

\def\paperID{*} 
\def\confName{ICCV}
\def\confYear{2025}

\title{Controllable Layer Decomposition for Reversible Multi-Layer Image Generation}

\author{
Zihao Liu$^1$\footnotemark[1],~ 
Zunnan Xu$^1$\footnotemark[1],~
Shi Shu$^1$,~
Jun Zhou$^1$\footnotemark[2],~
Ruicheng Zhang$^{1,2}$,~
Zhenchao Tang$^2$,~
Xiu Li$^1$\footnotemark[2]
\\
$^1$Tsinghua University,~
$^2$Sun Yat-sen University
}

\begin{document}

\twocolumn[{
\renewcommand\twocolumn[1][]{#1}

\maketitle
\begin{center}
    \centering
    \captionsetup{type=figure}
    \includegraphics[width=1\linewidth]{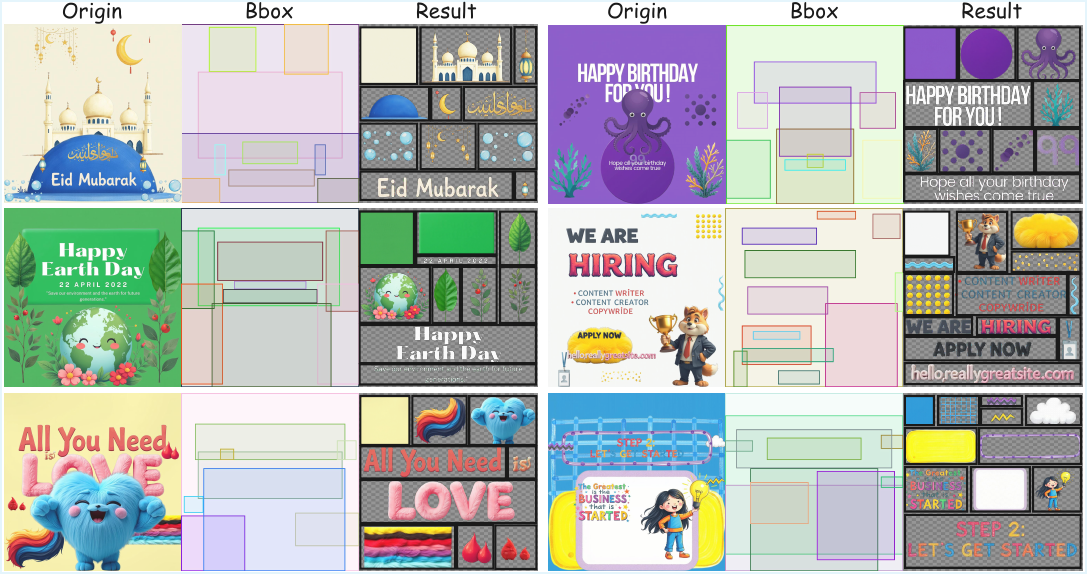}
    \caption{
    Given an image and its bounding boxes, our framework generates a clean foreground layer for each box and a single, coherent background. Our method can cleanly separate every foreground layer even when objects are crowded and heavily overlapping, while keeping boundaries sharp, depth order intact, and the composite visually coherent.
    }
    \label{fig:show}
\end{center}
}]
\renewcommand{\thefootnote}{\fnsymbol{footnote}}
\footnotetext[1]{Equal Contribution.} 
\footnotetext[2]{Corresponding author.}

\begin{abstract}
This work presents Controllable Layer Decomposition (CLD), a method for achieving fine-grained and controllable multi-layer separation of raster images. In practical workflows, designers typically generate and edit each RGBA layer independently before compositing them into a final raster image. However, this process is irreversible: once composited, layer-level editing is no longer possible. Existing methods commonly rely on image matting and inpainting, but remain limited in controllability and segmentation precision. To address these challenges, we propose two key modules: LayerDecompose-DiT (LD-DiT), which decouples image elements into distinct layers and enables fine-grained control; and Multi-Layer Conditional Adapter (MLCA), which injects target image information into multi-layer tokens to achieve precise conditional generation. To enable a comprehensive evaluation, we build a new benchmark and introduce tailored evaluation metrics. Experimental results show that CLD consistently outperforms existing methods in both decomposition quality and controllability. Furthermore, the separated layers produced by CLD can be directly manipulated in commonly used design tools such as PowerPoint, highlighting its practical value and applicability in real-world creative workflows. Our
project is available at \href{https://monkek123king.github.io/CLD_page/}{CLD}.
\end{abstract}
    
\section{Introduction}
\label{sec:intro}

In the design of posters, advertisements, and other visual media, designers typically do not directly work on a single raster image. Instead, they create and edit foreground elements across multiple RGBA layers. Specifically, designers use tools such as Adobe Photoshop or PowerPoint to construct foreground elements at the layer level and then overlay multiple foreground layers onto a background layer to produce the final image. However, this compositing process is irreversible: once multiple RGBA layers are merged into a single raster image, the original multi-layer information cannot be recovered. Consequently, when only a raster image is available, precise layer-level editing and adjustments become extremely challenging. Accurate layer decomposition from raster images can effectively address this issue, enabling designers to perform efficient and controllable edits on the separated layers.

In this work, we investigate the problem of user-controllable layer decomposition, aiming to split a single raster image into a set of independent layers based on user-provided bounding boxes. Existing approaches, such as decomposition methods for natural images~\cite{tan2016decomposing, akimoto2020fast, jin2024alignment, aksoy2017unmixing}, often suffer from incomplete foreground extraction or unwanted artifacts, including messy edges or background color leaking into foreground layers. Some studies~\cite{chen2025rethinking, tudosiu2024mulan, zhou2025fireedit} adopt a modular, multi-stage pipeline, dividing the task into object detection~\cite{redmon2016you, redmon2018yolov3, ren2015faster}, segmentation~\cite{xu2023bridging, kirillov2023segment, xu2024enhancing, huang2025densely, huang2025sam}, image matting~\cite{yu2021mask, sun2021semantic, kim2025zim}, and image inpainting~\cite{ju2024brushnet, ma2025followyourclick, corneanu2024latentpaint}, with each stage handled by a dedicated pre-trained model. However, such pipelines are prone to error propagation: mistakes in early stages can adversely affect subsequent stages, degrading overall layer decomposition quality. To address these limitations, LayerD~\cite{suzuki2025layerd} proposed a fully automatic graphic layer decomposition method, iteratively performing image matting for top-layer extraction and image inpainting for background completion. However, LayerD entirely relies on the matting model to identify top-layer elements, leaving users with no control over the final decomposition results.

To overcome these limitations, we propose a controllable layer decomposition method, which allows users to specify desired decomposition outcomes via bounding boxes. Our core model uses pre-trained DiT models (e.g., FLUX.1[dev]~\cite{blackforest2024flux}), fully exploiting their strong image generation capabilities to produce layer decomposition results of higher quality. Figure \ref{fig:show} shows the results of our method. It can be seen that the model is able to generate the corresponding foreground and background layers based on the provided bounding boxes, achieving excellent overall visual quality.
Moreover, previous evaluation metrics for layer decomposition were limited by uncontrollable generated results, lacking standardized benchmarks. By introducing bounding boxes, we can define ideal target layers and propose a new benchmark. Specifically, we establish a benchmark for controllable layer decomposition, built on the PrismLayersPro~\cite{chen2025prismlayers} dataset, with new metric dimensions tailored for controllable settings. Our main contributions are summarized as follows:

\begin{itemize}
\item {We propose CLD (Controllable Layer Decomposition), a controllable framework for raster image layer decomposition that leverages DiT’s generation and reasoning capabilities to achieve high-quality layer separation.}
\item {We introduce a new benchmark to assess the performance of layer decomposition from multiple dimensions.}
\item {Experimental results demonstrate that CLD outperforms baseline methods in both controllability and generation quality, and separated layers can be applied directly to downstream graphic design and editing tasks.}
\end{itemize}
\section{Related work}
\label{sec:related_work}

\noindent\textbf{Image Layer Decomposition.}
Image layer decomposition aims to divide an image into a set of composable layers that can be re-synthesized to reconstruct the original image. Early approaches are mostly color-based, grouping pixels by color similarity and focusing on digital painting or natural images~\cite{tan2016decomposing, akimoto2020fast, aksoy2017unmixing, tan2018efficient, koyama2018decomposing, horita2022fast}. Other studies explore object-level decomposition in natural scenes~\cite{isola2013scene, liu2024object, monnier2021unsupervised, zhang2025zero, zhan2020self, zheng2021visiting, ma2024followpose, xu2025hunyuanportrait, yang2025generative}. 
Recent studies have explored multi-layer image generation, using generative models to directly synthesize and disentangle foreground and background layers in a unified framework~\cite{huang2025dreamlayer, wang2025diffdecompose, qian2025layercomposer, pu2025art, chen2025prismlayers, zhang2024transparent, ma2024followyouremoji, huang2025psdiffusion, ma2025controllable}. For example, Huang et al.~\cite{huang2025dreamlayer} introduced DreamLayer, which explicitly models the interaction between transparent foreground and background layers for coherent text-driven generation. However, this method is designed for text-to-image synthesis and is not directly applicable to our layer decomposition task.
%
In design images such as posters and advertisements, multiple visual elements coexist, making decomposition challenging in terms of fine-grained separation and controllability. Recent works~\cite{chen2025rethinking, suzuki2025layerd} have explored this direction. Chen et al.\cite{chen2025rethinking} proposed a pipeline combining visual language models (VLM) with SAM\cite{kirillov2023segment}, while Suzuki et al.~\cite{suzuki2025layerd} introduced LayerD, which separates layers via image matting and inpainting. However, LayerD relies entirely on the matting model to determine top-layer elements, offering no user control and tending to produce coarse, element-level separations. To overcome these limitations, we propose a DiT-based framework that enables controllable, fine-grained layer generation guided by user-provided bounding boxes.

\noindent\textbf{Image matting and Element extraction.}
Natural image matting optimizes the compositing equation $I = \alpha F + (1-\alpha)B$ for $\alpha, F, B$ using inputs like trimaps or scribbles~\cite{xu2017deep, li2020natural, park2022matteformer}. Although trimap-free methods~\cite{ke2022modnet, yao2024vitmatte} reduce manual effort, they are optimized for photographic statistics and fail on the clean edges, solid colors, and repetitive textures common in graphic designs.

Zero-shot segmentation foundation models offer an alternative. SAM~\cite{kirillov2023segment} and SAM 2~\cite{ravi2024sam} produce binary masks from prompts but ignore partial transparency. HQ-SAM~\cite{ke2023segment} refines boundaries but remains at the instance level, failing to model soft transitions. ZIM~\cite{kim2025zim} adapts SAM for zero-shot matting, but it is trained for \emph{natural} objects and hallucinates alpha on solid-color graphics. Furthermore, none of these models support sequential decomposition; users cannot extract nested elements (e.g., ``segment the red subtitle") without providing a new manual trimap.
Our framework solves this. We treat the bounding box as a \emph{layer request} rather than a rough separator. The network predicts a hard binary mask (for crisp boundaries) and a residual alpha map (for optional soft transitions), enabling the extraction of both solid text and semi-transparent shadows. Because the model is conditioned on the box, it can be applied recursively: after removing the top element, the model can process the revealed region, yielding a full layer stack. Extensive experiments (Sec.~\ref{sec:experiment}) show our strategy outperforms matting-based and SAM-based baselines on graphic-design benchmarks in edge accuracy, layer consistency, and user controllability.
\section{Method}
\label{sec:method}

\begin{figure*}[ht]
\centering
\includegraphics[width=1.\textwidth]{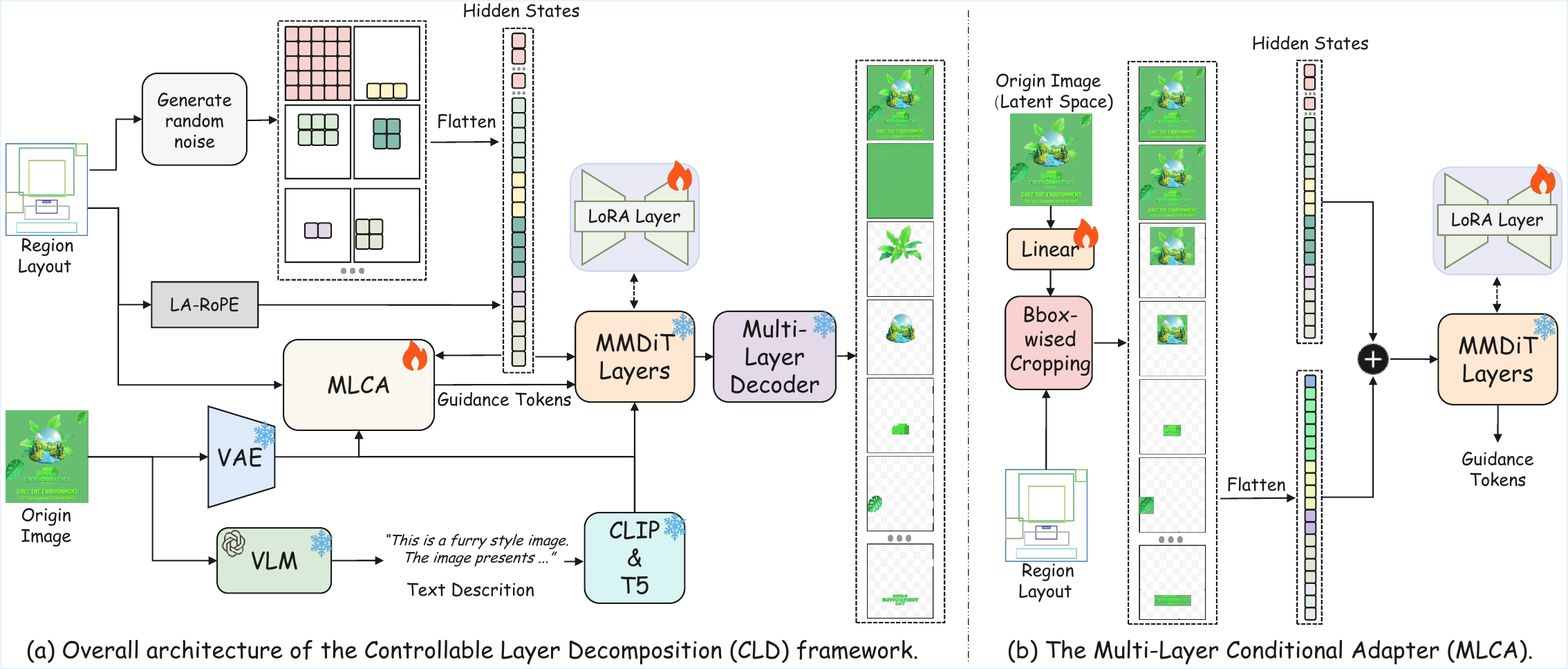}
\caption{Our framework utilizes a main backbone and a parallel control module for precise layer decomposition. 
\textbf{(a)} The overall CLD architecture, showing the LayerDecompose-DiT (LD-DiT) backbone responsible for generating the multi-layer latent. 
\textbf{(b)} The detailed structure of the Multi-Layer Conditional Adapter (MLCA). MLCA additively fuses features from the conditional image with the LD-DiT's hidden states, then performs hierarchical cropping based on the input bounding boxes to create a multi-layer guidance token sequence.
}
\label{fig:framework}
\end{figure*}

\begin{figure}
    \centering
    \includegraphics[width=\linewidth]{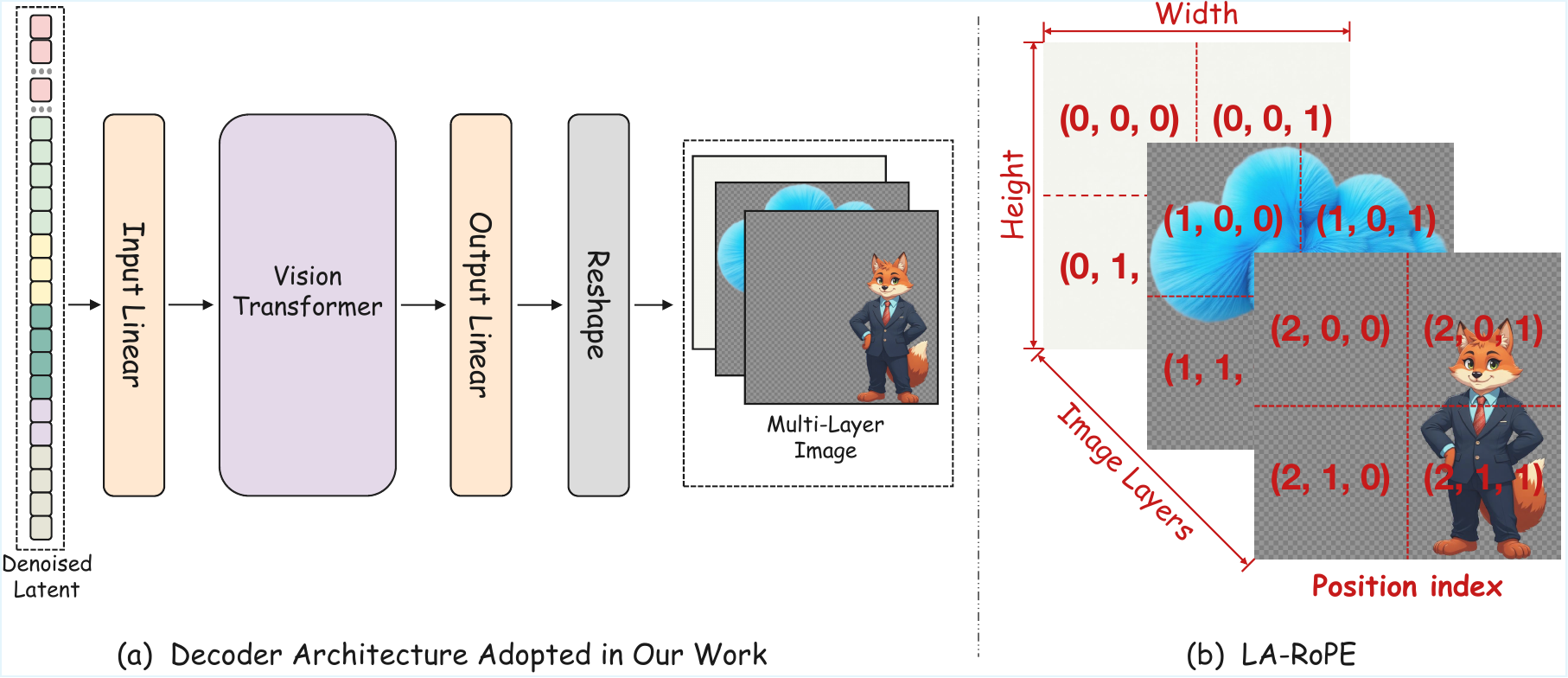}
    \caption{Overview of our adapted Multi-Layer RGBA image Decoder Architecture and Layer-Aware Rotary Position Encoding.}
    \label{fig:rope}
\end{figure}

\subsection{Preliminaries}
Flow Matching~\cite{lipmanflow, esser2024scaling} is an alternative to diffusion models by directly learning the velocity field that transports noise to data. Unlike noise-prediction diffusion models (e.g., DDPM~\cite{ho2020denoising}), it formulates generation as a continuous dynamical system, enabling a more stable and efficient sampling process.
%
Specifically, given a real sample $\mathbf{x}_0$ and a noise sample $\mathbf{x}1$, Flow Matching learns a velocity field $v_\theta(\cdot)$ to approximate the optimal transport between them. The objective aligns the predicted velocity with the true derivative $\frac{d\mathbf{x}_t}{dt}$ along the interpolated path $\mathbf{x}_t = (1-t)\mathbf{x}_0 + t\mathbf{x}_1$. By modeling the velocity rather than the noise residual, it achieves smoother optimization and more stable gradients in high-dimensional spaces. Due to its faster convergence and higher generation quality, we adopt Flow Matching as the training paradigm for CLD and fine-tune the backbone via LoRA~\cite{hu2022lora}. The overall training objective is:
\begin{equation}
    \mathcal{L}_{\text{FM}} = \mathbb{E}_{t \sim \mathcal{U}(0,1)}\left[
\left| \mathbf{v}_\theta(\mathbf{x}_t, t, c_{\text{text}}, c_{\text{img}}) - (\mathbf{x}_1 - \mathbf{x}_0) \right|^2 \right],
\end{equation}
where $c_{\text{text}}$ and $c_{\text{img}}$ denote the text and image conditions.

\subsection{Problem Formulation}

Given an raster image $\mathbf{I} \in \mathbb{R}^{H \times W \times 3}$, such as a composited design image, we regard it as consisting of a background layer $\mathbf{I}_{\mathrm{bg}} \in \mathbb{R}^{H \times W \times 4}$ and multiple foreground layers $\{\mathbf{I}_{\mathrm{fg}}^i \in \mathbb{R}^{H_i \times W_i \times 4}\}_{i=1}^{N-1}$, where $N$ denotes the total number of layers. Each layer is represented in RGBA format to preserve transparency information.
To enable controllable layer decomposition, the user can provide a set of bounding boxes $\mathbf{B}_u = \{B_1, …, B_{N-1}\}$, where each $B_i = (x_i^l, y_i^l, x_i^r, y_i^r)$ specifies the top-left and bottom-right coordinates of a target region. The background layer is defined by $B_0 = (0, 0, H, W)$, and together they form the complete bounding box set $\mathbf{B} = \text{concat}(B_0, \mathbf{B}_u)$.
Our objective is to generate a collection of $N$ disentangled layers $\mathbf{D} = \{ D_0, D_1, …, D_{N-1} \}$, where each $D_i$ represents an independent RGBA layer aligned with its corresponding bounding box $B_i$. The decomposition aims to ensure spatial alignment and visual consistency across all layers, thereby achieving fine-grained and controllable layer separation.

\subsection{LayerDecompose-DiT}

LayerDecompose-DiT (LD-DiT) is designed to simultaneously generate visual tokens for the background layer and multiple foreground layers, enabling fine-grained disentanglement and hierarchical modeling of image structures. At the architectural level, we build upon FLUX.1[dev]~\cite{blackforest2024flux}, which adopts the Multimodal Diffusion Transformer (MMDiT) as its backbone. MMDiT employs two distinct sets of network weights to separately process text tokens and image tokens, enabling efficient multimodal interaction. However, the original MMDiT is limited to single-image generation and cannot directly support multi-layer synthesis.
To address this limitation, we introduce several key enhancements. Specifically, the input visual tokens are cropped according to user-provided bounding boxes, and the target regional tokens are concatenated into a unified token sequence, which is then processed by MMDiT for denoising multi-layer generation. Moreover, to enable the model to better capture hierarchical dependencies within this sequence, we incorporate a Layer-Aware Rotary Position Embedding (LA-RoPE) to jointly encode spatial and inter-layer positional relationships. As shown in Figure \ref{fig:framework}, the conditional inputs to LD-DiT include the bounding boxes $\mathbf{B}$, the original image $\mathbf{I}$, and a text prompt $\mathbf{T}$. The text prompt can either be user-provided to semantically guide the generation process or automatically extracted from the original image using a VLM. The noisy inputs are obtained by adding Gaussian noise to the cropped multi-layer latent sequence, which is then denoised by MMDiT and decoded into a set of layered RGBA images.
%
To enhance global consistency in multi-layer generation, we introduce the composite image as an auxiliary generation target. Unlike layer-wise blending, it serves as a direct reconstruction of the input image. During training, the input image is embedded using the same visual encoder as the layer tokens and prepended to the layer-specific sequence, with a bounding box $(0, 0, H, W)$ covering the entire image. Formally, given layer tokens $\mathbf{X}_\text{layers} = \{\mathbf{x}_0, \mathbf{x}_1, \dots, \mathbf{x}_{N-1}\}$ and the composite token $\mathbf{x}_\text{comp}$, the input to MMDiT is:
\begin{equation}
  \mathbf{X}_\text{input} = [\mathbf{x}_\text{comp}; \mathbf{x}_0; \mathbf{x}_1; \dots; \mathbf{x}_{N-1}],  
\end{equation}
where $[;]$ denotes sequence concatenation. This setup enables the model to propagate global context through the composite token, reinforcing visual and structural coherence across layers.

\noindent\textbf{Latent Decoder.}
Conventional VAEs are limited to processing RGB inputs and are unable to capture transparency information. To address this limitation, our model explicitly incorporates transparency-aware encoding. We adapt two alternative designs into our framework: the Multi-Layer Transparent Image Autoencoder from ART~\cite{pu2025art} and the Latent Transparency module proposed in LayerDiffuse~\cite{zhang2024transparent}. Through systematic comparison, we find that the former achieves better preservation of alpha information and stronger inter-layer coherence. Based on these observations, we choose to adapt the Multi-Layer Transparent Image Autoencoder to fit our multi-layer generation pipeline, enabling LD-DiT to accurately maintain consistent structures across layers. Figure \ref{fig:rope} presents the decoder architecture used in our framework. 

\noindent\textbf{LA-RoPE.}
Traditional positional encodings typically model spatial relations within a single layer and fail to capture hierarchical dependencies across layers. To address this, we propose Layer-Aware Rotary Position Embedding (LA-RoPE), which jointly encodes both spatial and inter-layer positional information in a unified hierarchical form (see Fig.~\ref{fig:rope}). Each visual token is indexed by $[l, h, w]$, denoting its layer, height, and width positions. By integrating these hierarchical indices into the queries and keys of self-attention, LA-RoPE enables reasoning across layers and enhances structural coherence and controllability in multi-layer generation. Formally, let the $n$-th query and $m$-th key be $q_n, k_m \in \mathbb{R}^{d_\text{head}}$, each split along the channel dimension into three parts:
\begin{equation}
    q_n = \{q^l_n, q^h_n, q^w_n\}, \quad k_m = \{k^l_m, k^h_m, k^w_m\}.
\label{eq:pe_1}
\end{equation}
The (n,m)-th element of the attention matrix is:
\begin{equation}
    A(n,m) = \sum_{c \in \{l,h,w\}} \mathrm{Re}\big[ q^c_n (k^c_m)^* e^{i (p^c_n - p^c_m) \theta} \big],
\label{eq:pe_2}
\end{equation}
where $p_n = \{l_n, h_n, w_n\}$ denotes the 3D position index of the n-th token, $(k^c_m)^*$ is the complex conjugate of $k^c_m$, $\theta \in \mathbb{R}$ is a preset nonzero constant, and $\mathrm{Re}[\cdot]$ represents the real part of a complex number. 

\subsection{Multi-Layer Conditional Adapter}

Since the input to MMDiT consists of a multi-layer sequence of visual tokens, existing conditional control frameworks, such as ControlNet~\cite{zhang2023adding}, are primarily designed for single-layer image generation and therefore struggle to capture the complex dependencies among multiple visual layers. In particular, ControlNet performs conditioning through a global residual pathway, which offers only coarse-level guidance and lacks the ability to maintain structural alignment and inter-layer consistency.

To overcome these limitations, we introduce the Multi-Layer Conditional Adapter (MLCA), a dedicated conditioning module tailored for multi-layer visual generation. MLCA explicitly models hierarchical relationships between layers and dynamically injects layer-specific conditions into the diffusion process. By aligning conditional features with each layer’s token representation, MLCA enables precise, fine-grained control while preserving inter-layer coherence throughout generation.
Specifically, the input conditional image $\mathbf{I}$ is first encoded into its latent representation $\mathbf{z}_\text{img}$ via a VAE encoder $\mathcal{E}_{\text{VAE}}(\cdot)$:
\begin{equation}
    \mathbf{z}_\text{img} = \mathcal{E}_{\text{VAE}}(\mathbf{I}) \in \mathbb{R}^{H’ \times W’ \times C}.
\label{eq:vae}
\end{equation}
Next, a linear mapping network $\text{Linear}(\cdot)$ projects the latent feature into the same feature space as the hidden states $\mathbf{h} \in \mathbb{R}^{L \times D}$ of the main MMDiT model, obtaining $\hat{\mathbf{z}}_{\text{img}}$:
\begin{equation}
   \hat{\mathbf{z}}_{\text{img}} = \text{Linear}(\mathbf{z}_\text{img}) \in \mathbb{R}^{H’ \times W’ \times D}.
\label{eq:linear}
\end{equation}
Subsequently, we crop $\hat{\mathbf{z}}_{\text{img}}$ according to the input bounding boxes $\mathbf{B}$, slicing the corresponding regions and flattening them into a multi-layer guidance token sequence $\mathbf{h}_\text{img}$. This sequence is structurally aligned with the MMDiT input, enabling parallel modeling across layers within the Transformer’s self-attention mechanism for both explicit alignment and implicit coordination:
\begin{equation}
    \mathbf{h}_\text{img} = \text{Flatten}(\text{Crop}(\hat{\mathbf{z}}_\text{img}, \textbf{B})) \in \mathbb{R}^{L \times D}.
\label{eq:crop_flatten}
\end{equation}
We then perform additive conditioning by summing the mapped conditional features with the MMDiT hidden states to obtain the fused feature representation $\hat{\mathbf{h}}$:
\begin{equation}
   \hat{\mathbf{h}} = \mathbf{h} + \mathbf{h}_\text{img}. 
\label{eq:add_h}
\end{equation}
Afterward, $\hat{\mathbf{h}}$ is fed into LD-DiT as a crucial guidance signal during the denoising process. This design allows the Transformer to dynamically coordinate inter-layer interactions while maintaining the individual objectives of each layer. By decoupling layer-wise conditioning and enabling effective cross-layer communication, MLCA delivers precise control, stronger inter-layer coherence, and more consistent multi-layer decomposition compared to traditional single-layer control frameworks.

\subsection{Dual-Condition Classifier-Free Guidance}

Building on the fine-grained guidance from MLCA, we further introduce a dual-condition Classifier-Free Guidance (CFG) strategy to improve inter-layer consistency and semantic controllability. The model takes two conditional inputs: a text description $c_\text{text}$ and a reference image $c_\text{img}$. During inference, the process splits into a conditional and an unconditional branch. 
Unlike conventional CFG that discards all conditions in the unconditional branch, we retain $c_\text{img}$ while nullifying $c_\text{text}$.

The reason for retaining $c_\text{img}$ in the unconditional branch is twofold. (1) It provides a structural anchor, ensuring that both conditional and unconditional branches remain aligned in terms of spatial layout and hierarchical structure. (2) It enables semantic incremental separation: by subtracting the unconditional prediction from the conditional one, shared structural information is canceled, isolating the text-driven semantic contribution. 

The predicted flow velocity $\hat{v}$ is computed as:
\begin{equation}
\begin{aligned}
    \hat{v} =\;& v_\theta(x_t, t, \emptyset, c_\text{img})  \\
    &+ s \cdot \Big( 
        v_\theta(x_t, t, c_\text{text}, c_\text{img}) 
        - v_\theta(x_t, t, \emptyset, c_\text{img})
      \Big),
\end{aligned}
\label{eq:cfg}
\end{equation}
where $s$ denotes the CFG scaling factor, and $v_\theta$ is the flow velocity predicted by the model under different conditions. 

\section{Experiment}
\label{sec:experiment}

\begin{table*}[htbp]
 \caption{Comparison with LayerD~\cite{suzuki2025layerd} on the Crello~\cite{yamaguchi2021canvasvae} dataset test set. LayerD Metrics refers to the evaluation metrics proposed in LayerD, while Q-Insight Metrics denotes the evaluation metrics we introduce based on the Q-Insight~\cite{li2025q} model.}
\centering 
\resizebox{1.0\linewidth}{!}{ 
\begin{tabular}{l|ccc|ccc|ccc}
\toprule
\multirow{2}{*}{\textbf{Method}} & \multicolumn{3}{c|}{\textbf{LayerD Metrics}}& \multicolumn{3}{c|}{\textbf{Q-Insight Metrics}} & \multicolumn{3}{c}{\textbf{User Study}} \\ \cline{2-4} \cline{5-7} \cline{8-10}
\multicolumn{1}{c|}{} & \textbf{RGB L1$\downarrow$} & \textbf{\shortstack{Alpha \\ Soft IoU$\uparrow$}} & \textbf{\shortstack{Unified \\ Score$\downarrow$}} & \textbf{\shortstack{Semantic \\ Consistency$\uparrow$}} & \textbf{\shortstack{Visual \\ Fidelity$\uparrow$}} & \textbf{Editability$\uparrow$} & \textbf{\shortstack{Content \\ Completeness$\uparrow$}} & \textbf{\shortstack{Semantic \\ Consistency$\uparrow$}} & \textbf{\shortstack{Visual \\ Quality$\uparrow$}} \\ 
\midrule 
$\text{LayerD}_{[\text{ICCV 25}]}$ & 0.0653 & 0.7055 & 0.1799 & 3.8658 & 3.6773 & 4.0011 & 19\% & 24\% & 9\% \\
Ours & \textbf{0.0474} & \textbf{0.7771} & \textbf{0.1352} & \textbf{3.9157} & \textbf{3.7334} & \textbf{4.0462} & \textbf{81\%} & \textbf{76\%} & \textbf{91\%} \\
\bottomrule
\end{tabular}%
}
\label{tab:layerd}
\end{table*}

\begin{table*}[htbp]
\caption{Ablation study on different model variants.
We examine four configurations: (1) using the decoder adapted from LayerDiffuse~\cite{zhang2024transparent}, (2) disabling the image condition in the CFG unconditional branch, (3) removing the composite image prediction objective, and (4) the full model. 
The ART~\cite{pu2025art} adapted decoder is used for all experiments except the first configuration.
}
\center
\resizebox{1.0\linewidth}{!}{
\begin{tabular}{l|ccc|cc|ccc}
\toprule
\multirow{2}{*}{\textbf{Variant}} & \multicolumn{3}{c|}{\textbf{Layer-level}}& \multicolumn{2}{c|}{\textbf{Mask-level}} & \multicolumn{3}{c}{\textbf{Reconstruction}} \\ \cline{2-9} 
 \multicolumn{1}{c|}{}  & \textbf{PSNR$\uparrow$} & \textbf{SSIM$\uparrow$} & \textbf{FID$\downarrow$} & \textbf{IoU$\uparrow$} & \textbf{F1$\uparrow$} & \textbf{PSNR$\uparrow$} & \textbf{SSIM$\uparrow$} & \textbf{FID$\downarrow$} \\ \midrule
(1) Using adapted LayerDiffuse decoder
& 24.429 & 0.822 & 24.016 & 0.847 & 0.902 & 22.570 & 0.824 & 36.127 \\ 
(2) w/o image condition in CFG-Uncond & 21.691 & 0.767 & 69.910 & 0.576 & 0.692 & 20.658 & 0.819 & 103.250 \\ 
(3) w/o composite image prediction & 26.211 & 0.845 & 21.453 & 0.867 & 0.918 & 27.240 & 0.926 & 15.638 \\ 
(4) Full Model & \textbf{27.646} & \textbf{0.874} & \textbf{19.413} & \textbf{0.867} & \textbf{0.920} & \textbf{29.825} & \textbf{0.945} & \textbf{11.464} \\ 
\bottomrule
\end{tabular}%
}
\label{tab:ablation}
\end{table*}


\begin{figure*}
    \centering
    \includegraphics[width=1\linewidth]{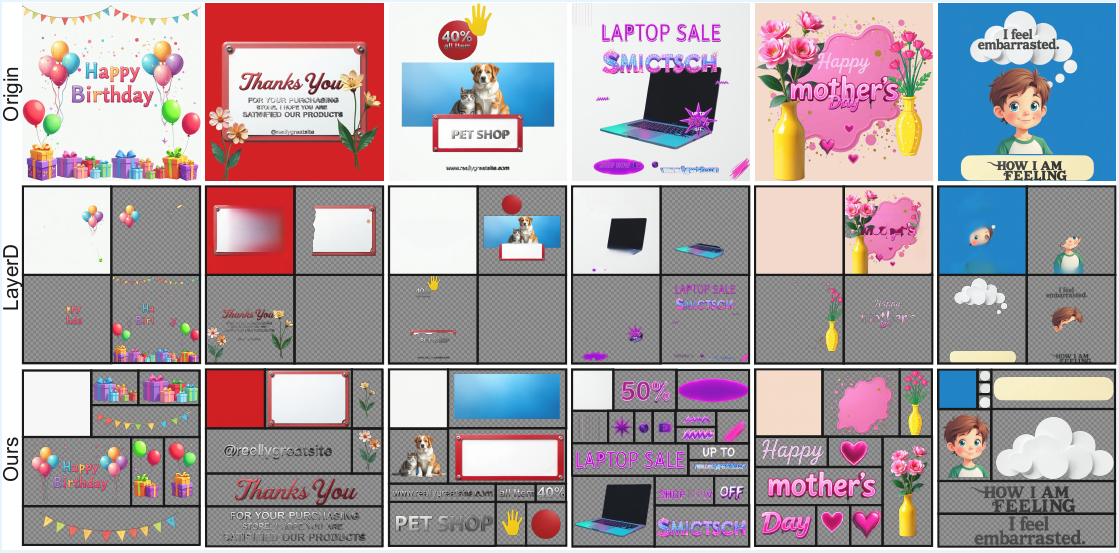}
    \caption{
    Unlike LayerD~\cite{suzuki2025layerd}, which offers  coarse separation and lacks user control, our method uses bounding boxes to guide a more fine-grained and controllable process. This results in better precision, visual quality, and hierarchical consistency in complex scenarios.
    }
    \label{fig:layerd}
\end{figure*}

\begin{figure}
    \centering
    \includegraphics[width=1\linewidth]{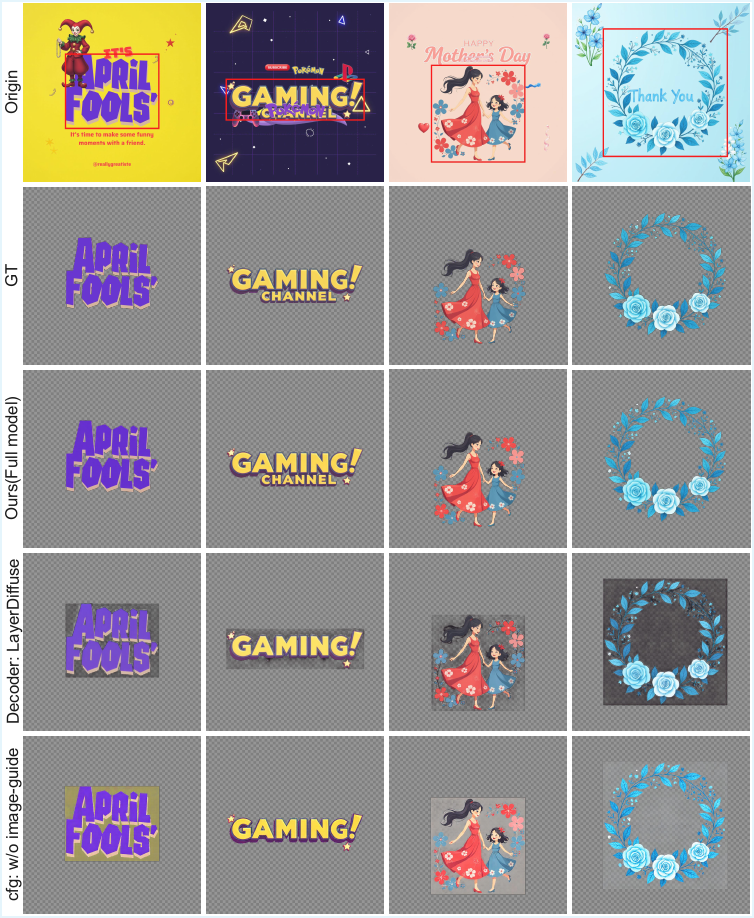}
    \caption{
    Ablation study on the impact of decoder choices and the CFG unconditional image condition.
    }
    \label{fig:ablation}
\end{figure}

\begin{figure}
    \centering
    \includegraphics[width=1\linewidth]{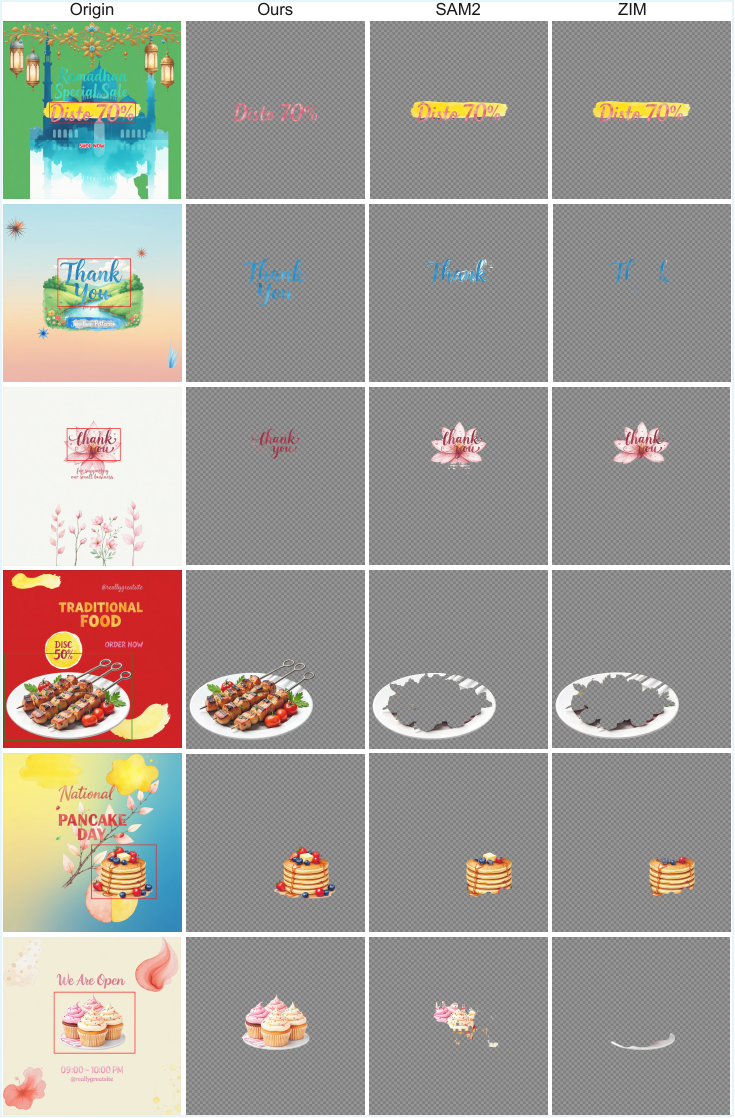}
    \caption{
    Unlike segmentation (SAM2~\cite{ravi2024sam}) and matting (ZIM~\cite{kim2025zim}) models, which fail on complex design images, our generative approach produces clearer, more coherent layering.
    }
    \label{fig:sam}
\end{figure}

\begin{figure}
    \centering
    \includegraphics[width=1\linewidth]{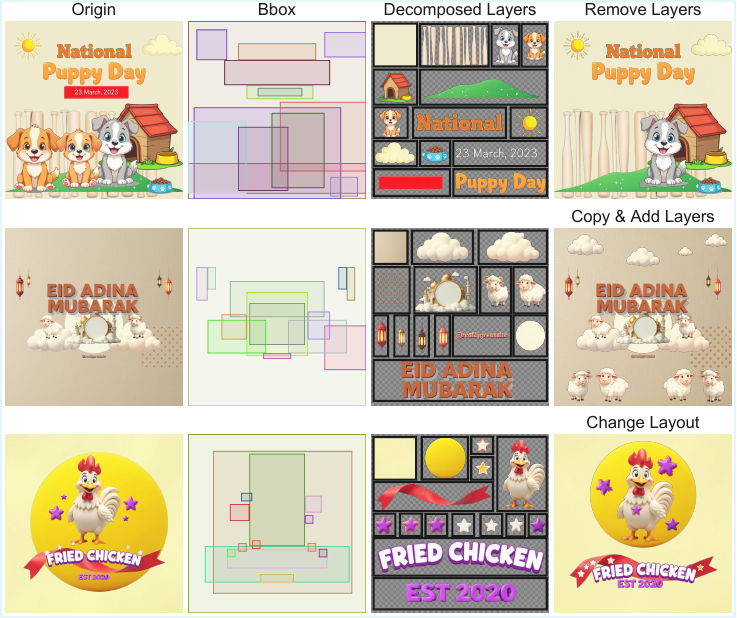}
    \caption{
    Application: Direct Editing on Decomposed Layers. This example showcases direct editing operations on separated layers, including Remove Layers, Copy \& Add Layers, and Change Layout. All operations are performed within PowerPoint.
    }
    \label{fig:edit}
\end{figure}

\subsection{Experiment Setting}

\noindent\textbf{Dataset Preparation.}
We employ PrismLayersPro~\cite{chen2025prismlayers}, one of the latest and largest high-quality multi-layer image datasets, as the primary source for both training and evaluation. This dataset contains approximately 20K groups of high-quality layered samples, where each group consists of a complete composite image, multiple corresponding transparent layers (in RGBA format), and an associated textual description. PrismLayersPro spans 21 distinct categories, encompassing a wide range of visual styles such as 3D, cartoon, and others, exhibiting high diversity and visual complexity. For data partitioning, we divide PrismLayersPro into training, validation, and testing sets with a ratio of 90\% / 5\% / 5\%, respectively. Based on this partitioning, we define a set of dedicated evaluation metrics and construct a new benchmark for multi-layer decomposition.

\noindent\textbf{Implementation details.}
Our backbone is built upon the latest FLUX.1[dev] model, a state-of-the-art Multimodal Diffusion Transformer (MMDiT) architecture. We fine-tune this model on the partitioned PrismLayersPro training set using LoRA adaptation, with the LoRA rank set to 64. The training is performed with the Prodigy optimizer, a learning rate of 1, and a batch size of 4. The model is trained for 25K iterations on images with a resolution of 1024 × 1024.

\subsection{Metrics}
\label{sec:exp_metrics}

Since there is currently no established benchmark for evaluating deterministic multi-layer image separation tasks, we propose a comprehensive, multi-dimensional evaluation to assess model performance across layer quality, transparency accuracy, and overall reconstruction fidelity. Our evaluation system is divided into three main categories:

\begin{itemize}
    \item \textbf{Layer-level}: These metrics assess the visual quality of each individual layer, using PSNR and SSIM for pixel-level and structural consistency, and FID~\cite{dowson1982frechet} for perceptual similarity between generated and real images.

    \item \textbf{Mask-level}: We use IoU (Intersection over Union) and F1 score to evaluate the alignment between generated and ground-truth alpha masks, reflecting accuracy in modeling transparent regions and boundaries.

    \item \textbf{Reconstruction}: All generated RGBA layers are composited back into an RGB image. We then evaluate this reconstructed image with PSNR, SSIM, and FID to assess global coherence and overall fidelity restoration.
\end{itemize}

All generated images are in RGBA format, while evaluation metrics (PSNR, SSIM, FID) are defined only for RGB images. To ensure metric compatibility and fair comparison, we first convert the RGBA image into RGB format. Specifically, We use a fixed neutral gray background $\mathbf{I}_{\mathrm{grey}}$ with pixel values $(0.5,0.5,0.5)$ within the range $[0, 1]$, and then composite it into $\mathbf{I}_{\mathrm{tgt}}$ with the RGBA image’s RGB channels $\mathbf{I}_{\mathrm{rgb}}$ and alpha channel $\mathbf{I}_{\alpha}$ as follows:

\begin{equation}
\mathbf{I}_{\mathrm{tgt}} = \mathbf{I}_{\mathrm{rgb}} \times \mathbf{I}_{\alpha} + \mathbf{I}_{\mathrm{grey}} \times (1 - \mathbf{I}_{\alpha}).
\label{eq:rgba2rgb}
\end{equation}
%
The use of a neutral gray background ensures robust blending with minimal color distortion, preserving transparent regions and enabling consistent evaluation across methods.

\subsection{Quantitative Result}
Currently, there is no existing work that performs a fully comparable multi-layer generative decomposition under user-guided conditions. Therefore, we select LayerD~\cite{suzuki2025layerd}, a state-of-the-art method for automatic layer separation, as our primary baseline for comparison since it is the most recent and directly comparable approach in both task definition and output representation.
%
LayerD~\cite{suzuki2025layerd} performs layer separation using image matting and inpainting. Operating on a single composite image, it relies entirely on the model’s inference, offering no explicit control over layer structure. To address layer correspondence, LayerD employs order-aware alignment via DTW~\cite{muller2007information} and computes evaluation metrics based on the matched layers.
It is important to note that our method differs from LayerD in that it incorporates user-provided bounding boxes to guide the layer decomposition. This bounding-box guidance introduces explicit structural control over the generated layers, which LayerD does not utilize.  Consequently, the two methods may produce different generation styles due to the additional structural guidance in our approach. We perform this comparison to demonstrate how the combination of bounding-box guidance and the proposed model architecture affects controllability, precision, and visual quality.

Table~\ref{tab:layerd} presents a comparison between LayerD and our proposed method under the evaluation metrics defined by LayerD, using the Crello~\cite{yamaguchi2021canvasvae} dataset as the test set, which we adopt for fair comparison since it was also used in LayerD's experiments. To adapt to our task, the RGB\_L1 metric is calculated by first converting RGBA images into RGB format according to Equation~\ref{eq:rgba2rgb}, and then applying the L1 distance calculation. The Unified\_Score metric jointly considers the performance of both RGB\_L1 and Alpha\_soft\_IoU, and is formulated as follows:
\begin{equation}
\text{Unified\_Score} = \frac{\text{RGB\_L1} + (1 - \text{Alpha\_soft\_IoU})}{2}.
\label{eq:unified_score}
\end{equation}
Experimental results show that our method achieves better results than LayerD across all evaluation metrics, benefiting from the introduction of bounding-box guidance and the powerful image generation capability of DiT.

To further assess our model’s performance, we employ Q-Insight~\cite{li2025q}, a state-of-the-art image quality evaluation model that leverages MLLM reasoning and reinforcement learning to provide interpretable, zero-shot perceptual quality assessment. We design three evaluation dimensions: Semantic Consistency, Visual Fidelity, and Editability. As shown in Table~\ref{tab:layerd}, our method achieves better performance than LayerD across all three dimensions, confirming its comprehensive advantages in semantic preservation, visual quality, and editability.

It is worth noting that the LayerD model is trained on the Crello~\cite{yamaguchi2021canvasvae} training set, while our model is trained solely on the PrismLayerPro~\cite{chen2025prismlayers} dataset. Despite this, our model still achieves better results on the Crello test set, further demonstrating its strong generalization ability and robustness across datasets.

\subsection{Qualitative Result}

\noindent\textbf{Comparison with LayerD.}
Figure~\ref{fig:layerd} shows a visual comparison of layer separation between our method and LayerD, and the results of the user study are presented in Table~\ref{tab:layerd}. Our model produces more detailed and fine-grained layer structures, benefiting from user-specified bounding boxes that guide the decomposition process. This additional guidance allows higher controllability and flexibility, particularly in complex scenes. Compared to LayerD’s automatic separation, our method preserves semantic and structural alignment across layers, and produces results that better align with user preferences.

\noindent\textbf{Comparison with Image Matting and Segmentation Methods.}
Recent layer separation methods, including our baseline LayerD~\cite{suzuki2025layerd}, largely rely on image matting or segmentation models, whose performance directly impacts subsequent layer separation and background inpainting. These models often struggle in scenarios containing with numerous foreground objects or complex hierarchical relationships, limiting the quality of the decomposition. To illustrate this, we compare our approach with widely used matting and segmentation methods, ZIM~\cite{kim2025zim} and SAM2~\cite{ravi2024sam}. Figure~\ref{fig:sam} shows that in scenes with overlapping elements, complex layouts, or text-rich regions, these methods tend to produce blurred boundaries, fragmented regions, or mis-segmented objects.
In contrast, our method performs multi-layer decomposition in a unified generative process. Leveraging the global reasoning and contextual modeling of DiT, it accurately infers layer boundaries and maintains visual and semantic coherence, even under heavy occlusion. The comparison results indicate that combining generative end-to-end decomposition with user-guided structural control provides inherent advantages over existing layer decomposition methods based on image matting or segmentation, offering higher fidelity, clearer layer separation, and stronger inter-layer consistency.



\subsection{Ablation Study}
\label{sec:ablation}

We conducted systematic ablation studies on our proposed layer decomposition model to evaluate the contribution of each component to the overall performance. The experiments were performed on a test set derived from the PrismLayersPro~\cite{chen2025prismlayers} dataset, and model performance was analyzed using the multi-dimensional evaluation metrics defined in Sec.~\ref{sec:exp_metrics}.

\noindent\textbf{Incorporating the composite image into the generation target.} 
In our model design, we include the composite image as an auxiliary generation target to enable information flow between the overall image and individual RGBA layers, improving global consistency and reconstructability. As shown in Table~\ref{tab:ablation}, this design enhances Layer-level and Mask-level metrics, with the largest gains in the Reconstruction metric, demonstrating improved visual coherence and layer fidelity.

\noindent\textbf{CFG unconditional branch.} 
Our framework uses two main conditions: a text description and the original image. During inference, we apply Classifier-Free Guidance (CFG), dropping the text condition in the unconditional branch but retaining the original image. Keeping the image in the unconditional branch significantly improves results (Table~\ref{tab:ablation}) and prevents background leakage in foreground layers (Figure~\ref{fig:ablation}), highlighting its importance for spatial consistency and semantic coherence.

\noindent\textbf{RGBA image decoder.} 
Traditional VAE decoders only handle RGB. We adapted two RGBA strategies for CLD: (1) LayerDiffuse~\cite{zhang2024transparent}, which reconstructs RGB and alpha separately, and (2) ART~\cite{pu2025art}, which directly generates RGBA from a ViT-based latent. As shown in Figure~\ref{fig:ablation} and Table~\ref{tab:ablation}, the ART-adapted decoder achieves better metrics and cleaner foreground boundaries.

\subsection{Application}
As shown in Figure~\ref{fig:edit}, we demonstrate a practical application of our method. By  manipulating the separated layers, users can conveniently perform various editing operations in common office software such as PowerPoint, including removing layers, copying and adding layers, and adjusting layouts. This layer-based editing paradigm greatly enhances the editability and flexibility of image content, enabling even non-professional users to easily accomplish complex element rearrangement and layout adjustments. This highlights the practical value of our method in real-world design and content creation scenarios.

\section{Conclusion}
\label{sec:conclusion}
In this paper, we present CLD, a method to separate raster images into fine-grained layers, solving a key limitation in post-production. We use bounding boxes for precise, user-guided separation and introduce a new multi-dimensional benchmark for this task. Experiments show CLD outperforms existing methods in decomposition quality, controllability, and visual fidelity. The generated layers are directly usable for downstream editing (e.g., removal, recomposition), offering a practical and high-quality solution for creative design and future multi-layer extensions.

{
    \small
    \bibliographystyle{ieeenat_fullname}
    \balance
    \bibliography{main}
}

\clearpage
\setcounter{page}{1}
\maketitlesupplementary

\section{Further Explanation of the Baseline}

\begin{figure*}
    \centering
    \includegraphics[width=\linewidth]{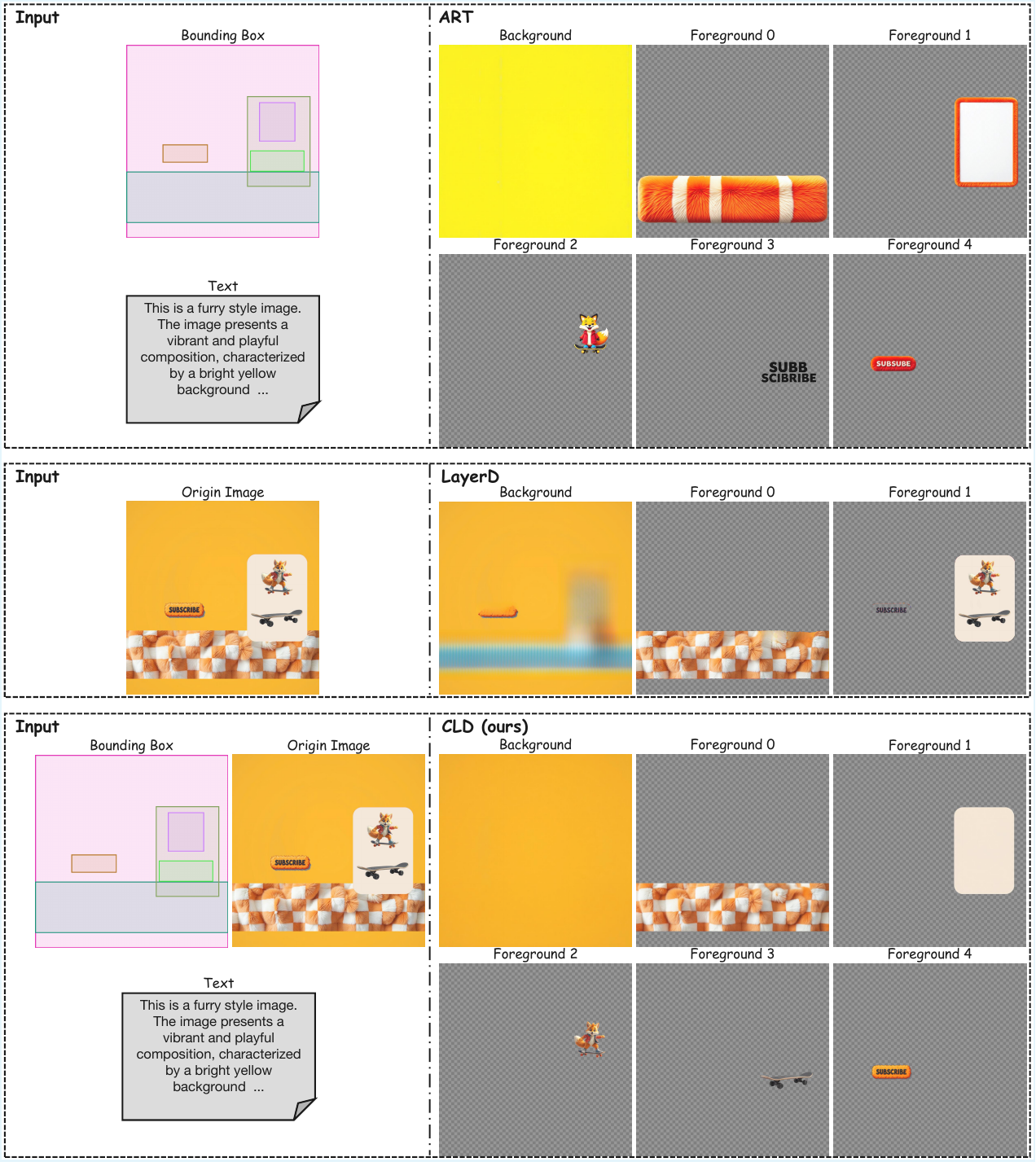}
    \caption{
        Comparison with related work. The figure illustrates the distinctions between our task and prior methods: ART~\cite{pu2025art} generates multilayer images without target-image constraints, and LayerD~\cite{suzuki2025layerd} decomposes layers without fine control. Our approach, by incorporating user-provided bounding boxes, achieves accurate and controllable layer separation.
    }
    \label{fig:baseline_comapre}
\end{figure*}

Currently, no existing work fully matches our task setting. Recent approaches~\cite{suzuki2025layerd, chen2025rethinking} are mostly based on image matting or vision-language models (VLMs), but these methods cannot handle user-specified bounding box constraints. Other related works~\cite{huang2025dreamlayer, pu2025art, zhang2024transparent} that use DiT focus on text-to-image generation rather than layer decomposition, making them unsuitable as direct baselines. Figure~\ref{fig:baseline_comapre} illustrates the differences between our task setup and those of related works.

Among existing research, the most relevant method is the recently proposed LayerD~\cite{suzuki2025layerd}. This work investigates three architectural variants: the matting-base model, the YOLO-base model, and the VLM-base model. Among them, the matting-base design achieves the strongest performance. In addition, LayerD systematically examines multiple combinations of matting models and inpainting models and selects the best-performing configuration. As a result, LayerD can be regarded as the current state of the art among traditional approaches based on matting/segmentation/VLM.

For these reasons, we adopt LayerD as our primary baseline. Experimental results demonstrate that our method significantly outperforms LayerD in both quantitative metrics and visual quality. This highlights the advantages of our diffusion-based layered generation framework, which provides stronger representational capacity and better disentanglement under complex scenarios compared to traditional matting/segmentation/VLM-based solutions.

\begin{figure}[ht]
    \centering
    \includegraphics[width=0.95\linewidth]{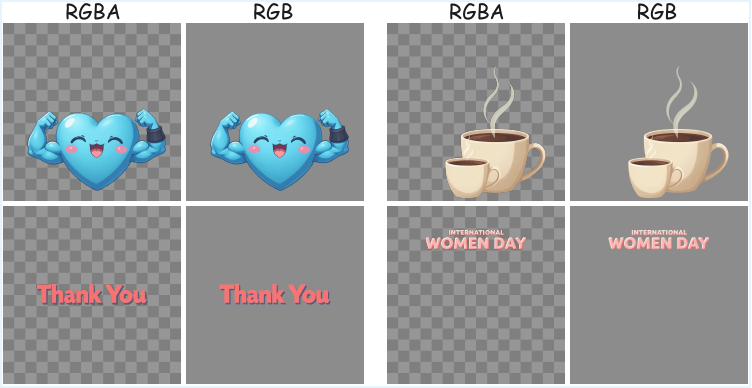}
    \caption{
        Visualization of RGBA–RGB Conversion.
    }
    \label{fig:rgba2rgb}
\end{figure}

\section{Q-Insight Additional Details}

To provide a more objective and comprehensive comparison of our method with LayerD~\cite{suzuki2025layerd} in terms of generation quality, we further employ Q-Insight~\cite{li2025q}, the current state-of-the-art image evaluation model, for automated assessment. We evaluate the generated images from multiple perspectives, complementing traditional metrics by assessing semantic alignment and editability. Specifically, we structure the assessment into three dimensions:

\begin{itemize}
    \item \textbf{Semantic Consistency}: Measures the alignment between the generated image and the textual description or target semantics;
    \item \textbf{Visual Fidelity}: Assesses visual quality, detail preservation, and overall realism;
    \item \textbf{Editability}: Evaluates the manipulability of the image for subsequent editing, modifications, and local adjustments.
\end{itemize}

The evaluation prompts used are as follows.

\begin{contentagnosticlayout}
You are evaluating the quality of a layered image decomposition. 

The input consists of multiple images: [Original Image], [Layer 1], [Layer 2], ..., [Reconstructed Image].

\# --- THINKING STEP ---

First, reason about the decomposition quality step-by-step inside the \texttt{<think>} tags. Analyze the semantic completeness of each layer, the visual fidelity of the reconstruction, and the practical editability of the layer structure based on the criteria below.

\# --- RATING STEP ---

Second, based on your reasoning, provide three separate ratings. 

All ratings should be floats between 1.00 and 5.00, rounded to two decimal places.

\# --- CRITERIA ---

1. **semantic\_consistency:** (1.00 = fragmented, meaningless layers; 5.00 = all layers are semantically whole and independent). 

Each layer should represent a complete, logical object or part (e.g., a whole text block, a distinct background element, a person).

2. **visual\_fidelity:** (1.00 = major artifacts, poor reconstruction; 5.00 = layers reconstruct the original perfectly). 

**Note: You must ignore the inherent artistic style of the image (e.g., photo vs. cartoon).**

Your score should *only* reflect artifacts from the decomposition process, such as halos, incorrect background inpainting, color bleeding, or missing pixels.

3. **editability:** (1.00 = useless, under- or over-decomposed; 5.00 = perfect granularity for editing).

**Preference: A finer-grained decomposition (more layers) is preferred and should receive a higher score,** as long as the individual layers still represent complete semantic parts.

Do not penalize for `over-decomposition' if the resulting fine-grained layers are logical and useful for an editor (e.g., separating a title from a body text is better than keeping them as one layer).

\# --- FORMATTING ---

Return the result in JSON format with the following keys:

\{

  ``semantic\_consistency": \texttt{<score>},
  
  ``visual\_fidelity": \texttt{<score>},
  
  ``editability": \texttt{<score>}
  
\}
)

\end{contentagnosticlayout}

\section{Additional Dataset Specifications}

The dataset used for training and in the proposed benchmark is based on PrismLayersPro, constructed by Chen et al\cite{chen2025prismlayers}. It contains 21 visual style categories, including 3D, Pokemon, anime, cartoon, doodle\_art, furry, ink, kid\_crayon\_drawing, line\_draw, melting\_gold, melting\_silver, metal\_textured, neon\_graffiti, papercut\_art, pixel\_art, pop\_art, sand\_painting, steampunk, toy, watercolor\_painting, and wood\_carving. The dataset comprises 20K samples, each providing complete layer information along with corresponding textual annotations. For each data sample, the dataset contains:

\begin{enumerate}
    \item A global textual description: $\mathbf{T}_{\text{global}}$;

    \item A composite image: $\mathbf{I}_{\mathrm{comp}}$;

    \item $n$ RGBA layers:
    \begin{equation}
      \{\mathbf{I}^{(0)}, \mathbf{I}^{(1)}, \dots, \mathbf{I}^{(n-1)}\},
    \end{equation}
    where $\mathbf{I}^{(0)}$ is the background layer, and $\mathbf{I}^{(1)} \sim \mathbf{I}^{(n-1)}$ are foreground layers;

    \item Text descriptions for each layer:
    \begin{equation}
        \{\mathbf{T}^{(0)}, \mathbf{T}^{(1)}, \dots, \mathbf{T}^{(n-1)}\};
    \end{equation}

    \item Bounding boxes for foreground layers: For the $k$-th foreground layer ($k \ge 1$), the bounding box is
    \begin{equation}
        \mathbf{b}^{(k)} = (x_{l}^{(k)},\, y_{l}^{(k)},\, x_{r}^{(k)},\, y_{r}^{(k)}),
    \end{equation}
    while the background layer does not provide a bounding box, defaulting to cover the entire image:
    \begin{equation}
        \mathbf{b}^{(0)} = (0, 0, H, W).
    \end{equation}

\end{enumerate}
 
PrismLayersPro provides sufficient data, complete layer annotations, and textual descriptions, allowing direct use for training and evaluation. We split each style category independently for our experiments: 0–90\% for training, 90–95\% for testing, and 95–100\% for validation, ensuring balanced and consistent partitions.

\section{RGBA-to-RGB Conversion Method}

We describe our RGBA-to-RGB conversion process  in Sec.~\ref{sec:exp_metrics}. A fixed solid background is used, and the input RGBA image is decomposed into RGB channels $\mathbf{I}_{\mathrm{rgb}}$ and the alpha channel $\mathbf{I}_{\alpha}$, with the final RGB image obtained via linear color compositing (Eq.~\ref{eq:rgba2rgb}). In this paper, we adopt neutral gray (pixel values (0.5, 0.5, 0.5) in the normalized range [0, 1]) as the default background color.

Since RGBA layers contain no “real background,” transparent regions must be filled with a placeholder color solely for computing standard RGB metrics (PSNR, SSIM, FID). The placeholder should satisfy two criteria: (1) it should not introduce additional color bias, and (2) it should minimize the evaluation error induced by background filling in a global statistical sense. Neutral gray, as the midpoint of the color space, naturally satisfies both criteria and provides a balanced fill, reducing systematic color shifts and worst-case error, leading to stable and fair evaluations.

From a visual perspective, neutral gray avoids artificial color or brightness shifts in transparent areas, ensuring comparability across methods. Moreover, in opaque regions (where $\alpha$ is close to 1), the influence of the background becomes negligible, and thus this choice does not interfere with foreground content. Figure~\ref{fig:rgba2rgb} illustrates the visual results after conversion.

In summary, using neutral gray (0.5, 0.5, 0.5) as a placeholder is statistically robust, visually neutral, and practically convenient, reducing metric bias and enabling consistent, reliable RGB-based comparisons.

\section{Additional Results}

\begin{figure*}
    \centering
    \includegraphics[width=\linewidth]{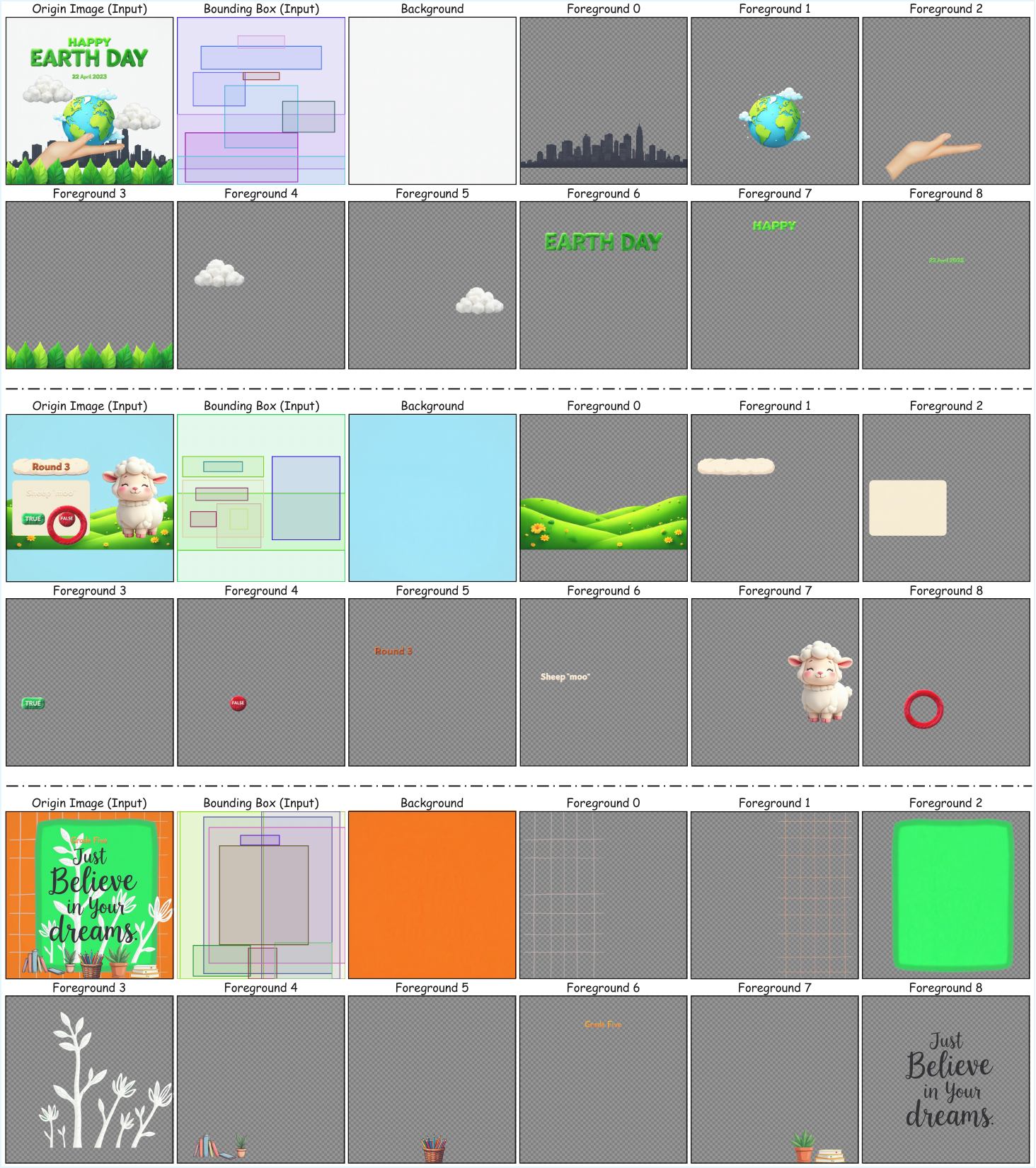}
    \caption{
        More qualitative results.
    }
    \label{fig:supp_quality_1}
\end{figure*}

\begin{figure*}
    \centering
    \includegraphics[width=\linewidth]{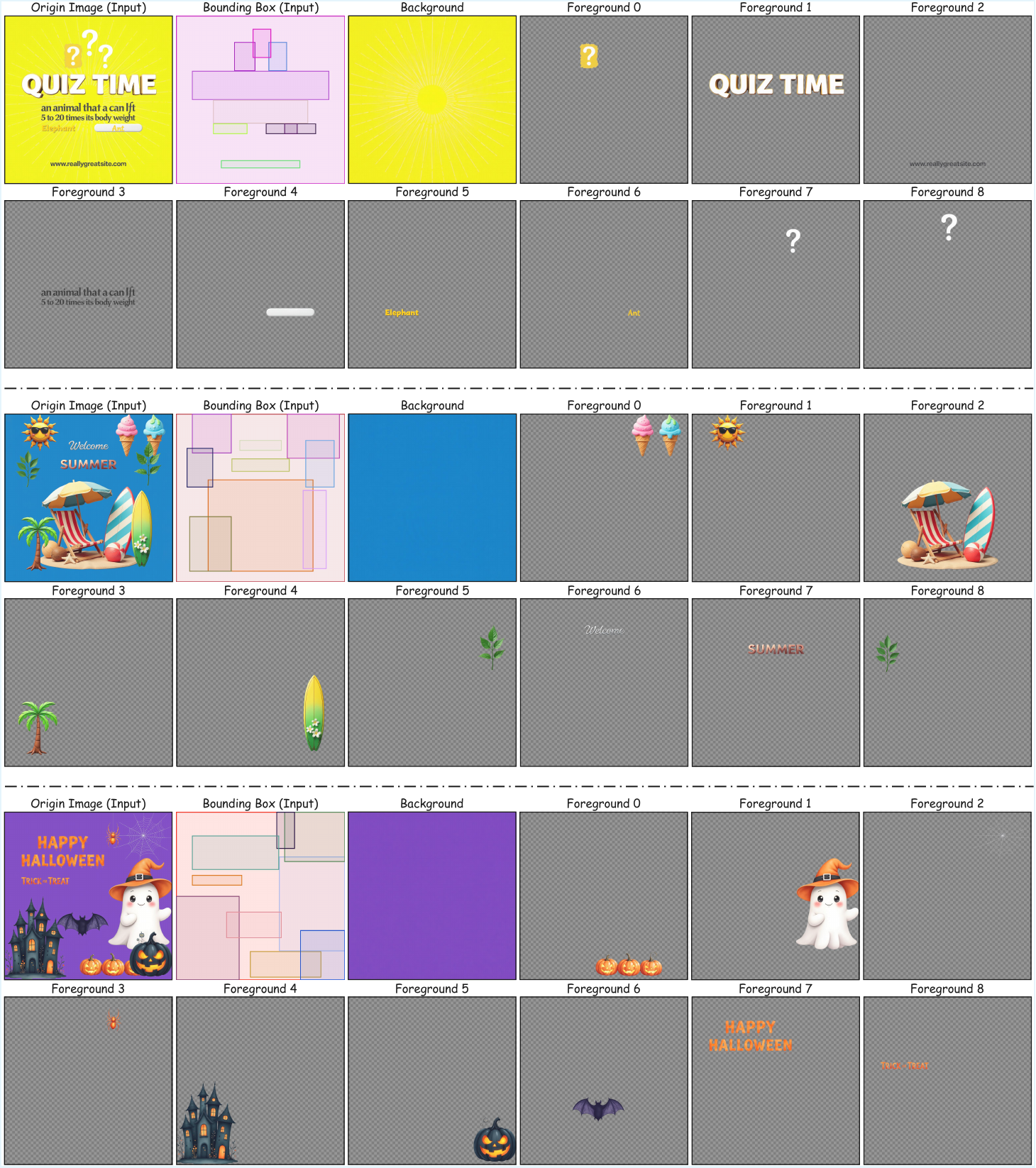}
    \caption{
        More qualitative results.
    }
    \label{fig:supp_quality_2}
\end{figure*}

In Figure~\ref{fig:supp_quality_1} and~\ref{fig:supp_quality_2}, we present additional examples of layer decomposition results. All samples shown in this figure are taken from the test split of our curated PrismLayersPro dataset. These examples further illustrate the effectiveness and consistency of our decomposition pipeline across diverse scenes and appearance variations.

\section{In-The-Wild Case}

\begin{figure*}
    \centering
    \includegraphics[width=\linewidth]{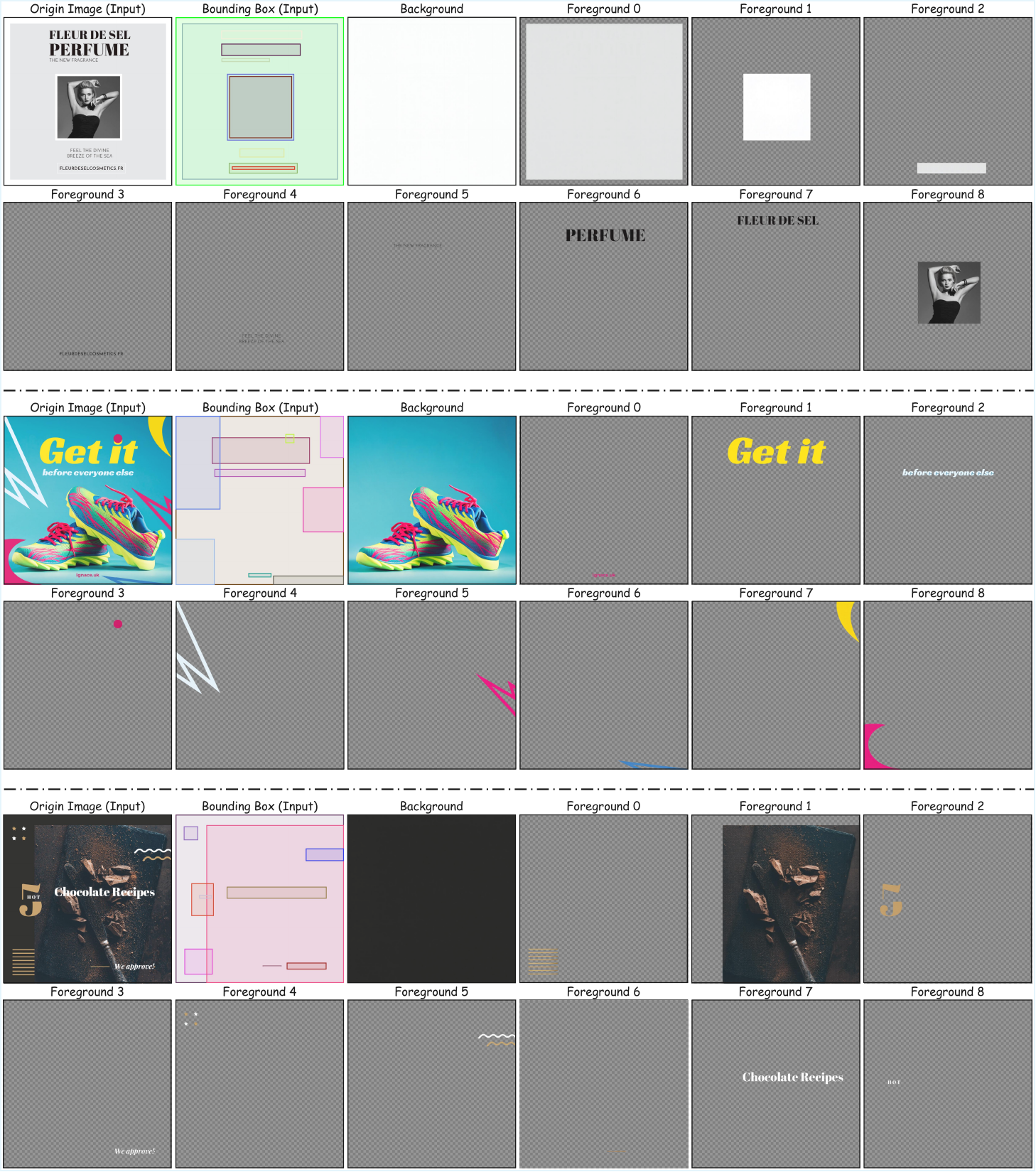}
    \caption{
        In-The-Wild Case.
    }
    \label{fig:supp_wild_1}
\end{figure*}

\begin{figure*}
    \centering
    \includegraphics[width=\linewidth]{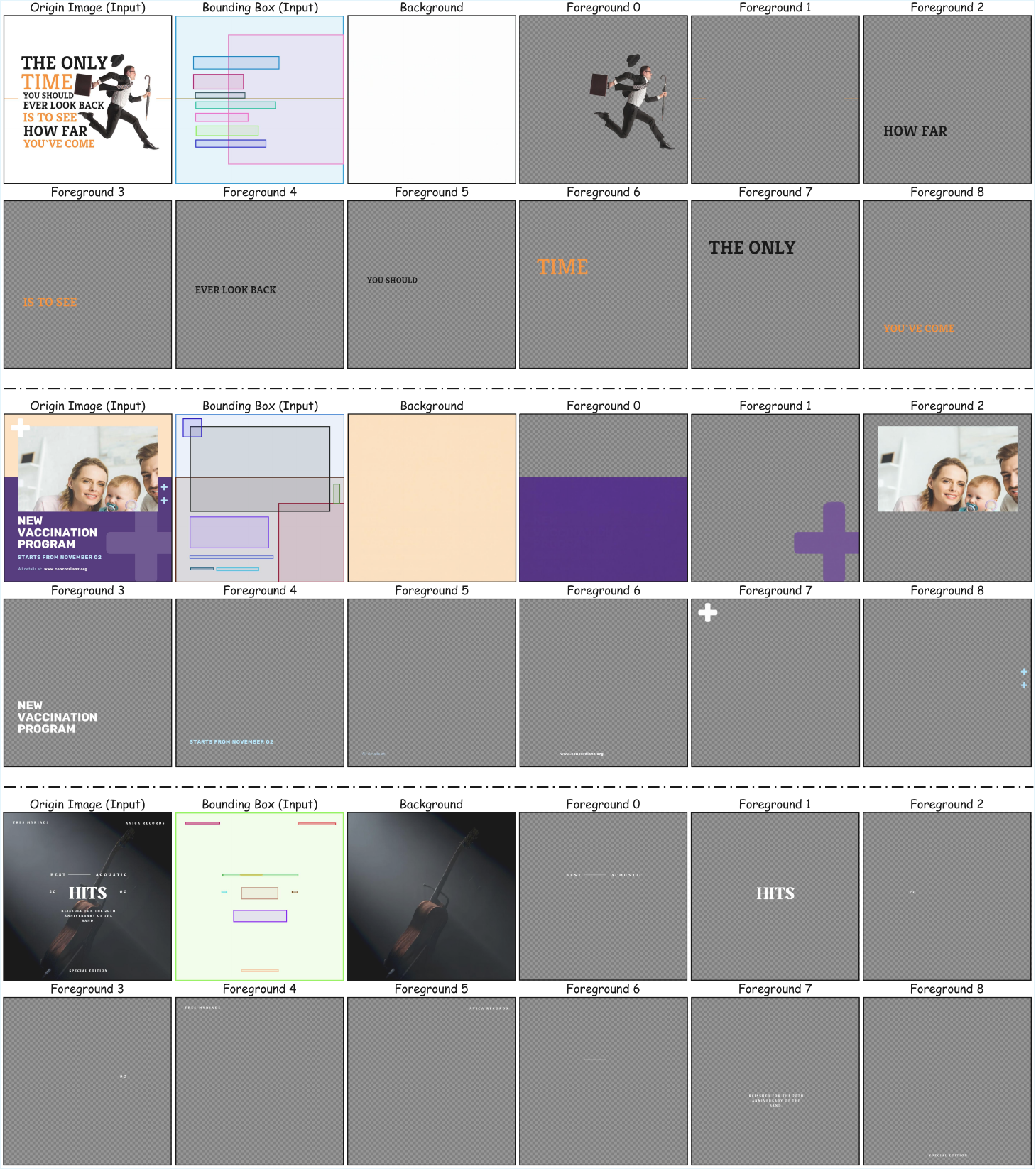}
    \caption{
        In-The-Wild Case.
    }
    \label{fig:supp_wild_2}
\end{figure*}

In Figure~\ref{fig:supp_wild_1} and~\ref{fig:supp_wild_2}, we present the layer decomposition results of our method on in-the-wild examples. These samples are drawn from the test split of the Crello~\cite{yamaguchi2021canvasvae} dataset, which contains diverse, professionally designed graphic layouts with rich stylistic variations. As shown in the figure, our model generalizes well beyond the controlled training distribution and is able to separate complex visual elements, including icons, text segments, and composite shapes. This demonstrates the robustness of our framework and its ability to handle real-world design assets encountered in practical editing scenarios.

\section{Failure Case}

\begin{figure*}
    \centering
    \includegraphics[width=\linewidth]{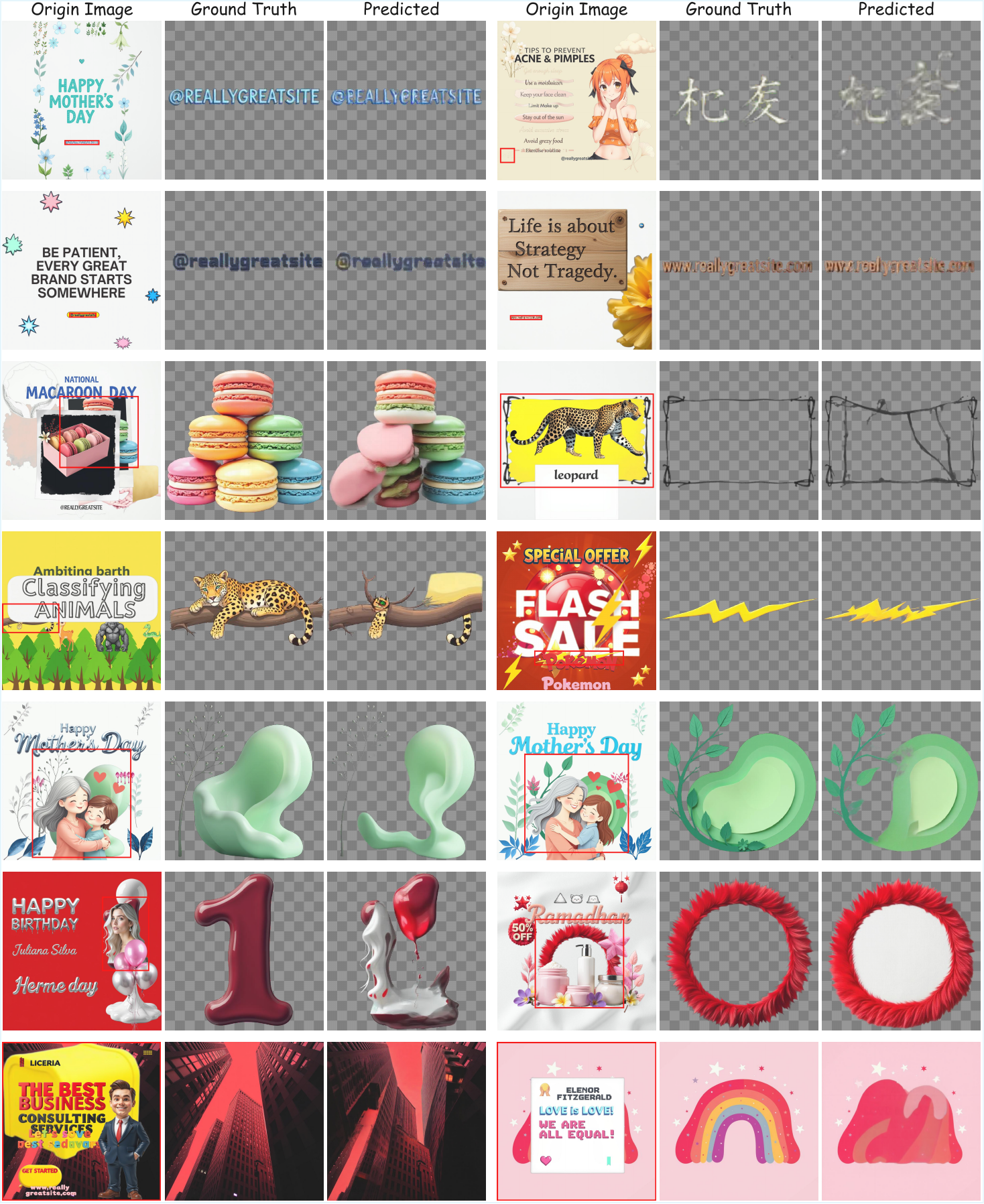}
    \caption{
        Failure case.
    }
    \label{fig:failure}
\end{figure*}

Figure~\ref{fig:failure} shows representative failure cases of our method, which fall into two main categories. 
The first occurs with very small, thin, or intricate content, such as fine text or tiny details, where limited spatial resolution prevents the model from capturing sufficient local cues. We believe that increasing input resolution or using multi-scale features could mitigate this issue. 
The second involves complex hierarchical occlusions. When objects are heavily occluded, the image provides insufficient information for accurate reconstruction. Although textual descriptions are used as auxiliary guidance, they are often insufficient to recover hidden content. Addressing this may require more advanced use of text cues or explicit structural priors.

\end{document}